\documentclass[twoside,fleqn]{article}
\usepackage{espcrc2}
\usepackage{psfig}
\usepackage{graphicx}

\newcommand{\cpt}{\mbox{$\chi$PT}}

\def\gsim{{\mathrel{\raise2pt\hbox to 8pt{\raise -5pt\hbox{$\sim$}\hss{$>$}}}}}
\def\rsim{{\mathrel{\raise2pt\hbox to 8pt{\raise -5pt\hbox{$\sim$}\hss{$>$}}}}}
\def\lsim{{\mathrel{\raise2pt\hbox to 8pt{\raise -5pt\hbox{$\sim$}\hss{$<$}}}}}

\def\etal{{\em et al.}}

\def\berlin#1{Nucl.\ Phys.\ {\bf B} (Proc.\ Suppl.) {\bf 106-107}, #1 (2002)}
\def\boston{these proceedings}
\def\prl#1{Phys.\ Rev.\ Lett.\ {\bf #1}}
\def\prd#1{Phys.\ Rev.\ {\bf D#1}}
\def\plb#1{Phys.\ Lett.\ {\bf #1B}}
\def\npb#1{Nucl.\ Phys.\ {\bf B#1}}

\newcommand{\lesssim}{\raisebox{-.6ex}{$\stackrel{\textstyle{<}}{\sim}$}}

\begin{document}

\title{
Panel discussion on chiral extrapolation of physical observables
\thanks{Edited by S.~Sharpe, with particular thanks to J.~Christensen 
for his detailed notes, which were essential for the reconstruction of the
responses and general discussion.}
}

\author{
Claude Bernard,\address{%
Department of Physics, Washington University,
St. Louis, MO 63130, USA}
Shoji Hashimoto,\address{%
High Energy Accelerator Research Organization (KEK),
Tsukuba, Ibaraki 305-0801, Japan}
Derek B.~Leinweber,\address{%
Centre for the Subatomic Structure of Matter and
Department of Physics and Mathematical Physics, University of
Adelaide, SA 5005, Australia}
Peter Lepage,\address{%
Physics Department, Cornell University, Ithaca, NY 14853, USA}
Elisabetta Pallante,\address{%
S.I.S.S.A., Via Beirut 2-4, 34014 Trieste, Italy}
Stephen R. Sharpe (chair)\address{%
Physics Department, Box 351560,
University of Washington, Seattle, WA 98195-1560, USA}
and Hartmut Wittig\address{%
      DESY, Theory Group,
      Notkestr. 85,
      D-22603 Hamburg,
      Germany}
}
      
\begin{abstract}
This is an approximate reconstruction of the panel discussion on 
chiral extrapolation of physical observables. The session consisted of
brief presentations from panelists, followed by responses from
the panel, and concluded with questions and comments from
the floor with answers from panelists. 
In the following, the panelists have summarized their statements,
and the ensuing discussion has been approximately reconstructed from notes.
%
\end{abstract}

\maketitle

\section{Introduction}

{\bf Sharpe:} It has become apparent 
from many talks at this conference 
that chiral extrapolation is an issue of great practical importance. 
Different approaches are being tried, and it is certainly timely to have a
general discussion of the issue. In order to focus the
discussion, I sent the panelists a draft list of key questions 
to focus their thoughts
as they were preparing their remarks. These questions
have evolved as a result of feedback, and my present version 
(in no particular order) is as follows.
\begin{enumerate}
\item
How small does the quark mass need to be to use chiral perturbation
theory ($\chi$PT)?
\item
Do we need to use fermions with exact chiral symmetry to reach the
region where $\chi$PT applies?
\item
What fit forms should we use outside the chiral region?
\item
Is the strange quark light enough to be in the chiral regime?
\item
Is it necessary to include $O(a,a^2)$ effects in the chiral Lagrangian?
\item
Can we use (present or future) partially quenched simulations to
obtain quantitative results for physical parameters?
\item
Can we use quenched simulations to give quantitative results
for physical parameters?
\item
Is it possible and/or desirable to work at $m_q=0$?
\end{enumerate}

\newpage
\section{Presentations}

{\bf Bernard:}
At this conference, and in the recent literature, several
groups have emphasized a key point about chiral extrapolations:
Over the typical current range of lattice values for the 
light quark masses, the data for many physical quantities is quite 
linear.  Yet linear extrapolations will miss the chiral 
logarithms that we know are present and therefore may introduce large 
systematic errors into the results.  JLQCD \cite{JLQCD-LAT01}, 
Kronfeld and Ryan \cite{KRONFELD}, and
Yamada's review talk here \cite{YAMADA} 
have stressed the relevance of this point for
heavy-light decay constants; while the Adelaide group 
\cite{TonyConf,Young:2002cj,Young:2001nc,Leinweber:1999ig,Leinweber:2001ac}
has brought out the
same point in the context of baryon physics.  All these groups deserve
a lot of credit for bringing this important issue to the fore.

Now the question is: ``What are we going to do about it?''
Attempts to extract the logarithms directly in the current typical 
mass range are in my opinion  doomed to failure: The extreme
linearity of the data indicates, at best, that higher order terms must
be contributing in addition to the logarithms, or, at worst, that we
are out of the chiral regime altogether.  The only real solution is
to go to lower quark masses.  We need to be well into the chiral regime,
to see the logs and make controlled fits including this known chiral
physics.  My rough guess in the heavy-light decay constant case is
that we need $m_\pi/m_\rho\approx0.3$, or at best $m_\pi/m_\rho \lsim 0.4$.
The latter range may be reachable, with significant work, with Wilson-type
fermions; while the former may require, at least
in the near term, staggered fermions.
The use of staggered light valence quarks in heavy-light simulations, 
as was suggested by Wingate at {\it Lattice 2001} \cite{WINGATE},  should make the 
chiral regime for that problem accessible very soon.

A different approach has been advocated by the Adelaide group.  They say
we can take into account the chiral logarithms in the current range of masses
by modeling the turn-off of chiral logarithms with a quantity-dependent
cutoff that represents the ``core'' of the object under study.  
I have nothing against modeling {\it per se}; I think it can be
an excellent tool to gain qualitative insight into the physics.  What I
think is wrong, or at least wrong-headed, about
the Adelaide approach is the suggestion that one can use it
to extract reliable quantitative answers with controlled errors.  Extraction
of such answers is after all why we are doing lattice physics in the first
place.  

The Adelaide model introduces a single parameter, the core size, to describe
the very complicated real physics involving couplings to all kinds
of particles --- $\rho$'s, $\sigma$'s, {\it etc.} --- as one moves
out of the chiral regime.  
The change in their results when they change the parameter by some amount or
vary the functional form at the cutoff is simply not a reliable,
systematically improvable error.   In other words, their
model is an uncontrolled approximation.

Suppose, however, one phrases the question in the following way: 
``Given some lattice data in the linear regime, are you likely to get
closer to the right answer with a linear fit, or with an Adelaide
form that interpolates between linear behavior and the known chiral
behavior at low mass?''  Phrased that way, my answer would be,
``the Adelaide form.''  But the problem is that, while you are most
likely closer to the right answer, you do not know the 
size of the errors --- unless you know the right answer to
begin with!  In my opinion, the linear fit is a ``straw man'' alternative.
The real alternative is to go to lighter masses and fit to the known
chiral form.  This approach, and this approach only, will produce
controlled, systematically improvable errors:  To improve, just go to
higher order in the chiral expansion or to still lighter masses.

Now if we want to go to lighter masses, I would argue that the
easiest way to do so is by using staggered fermions.
Dynamical staggered fermions
are very fast, and they have an exact lattice chiral symmetry.
However, as you know, many of the other staggered
symmetries are broken at finite lattice spacing.  

First of all, let me talk about nomenclature.
I would like to advocate here the use of the word ``taste'' to describe 
the 4 internal
fermion types inherent in a single staggered field.  Taste symmetry
is violated on the lattice at ${\cal O}(a^2)$ but becomes exact in 
the continuum limit.
I reserve the word ``flavor'' for different staggered fields, which
have an exact lattice symmetry (in the equal mass case) that mixes them.
For example, MILC is doing simulations with 3 flavors ($u$, $d$, and $s$) 
with $m_u=m_d\not=m_s$. Normally each flavor would have 4 tastes,
but we do the usual trick of taking the fourth root of the 
determinant to get a single taste per flavor.  Of course this is ugly
and non-local, and one must test that there are no problems introduced in the
continuum limit.

I find it useful to think about the effects of taste symmetry breaking as
just a more complicated version of ``partial quenching''  \cite{CB-MG-PQQCD}.
Sharpe and Shoresh \cite{SHARPE-SHORESH} have taught us that, as long as a theory has the right
number of sea quarks (3), the chiral parameters are physical
even if the masses of the quarks are not physical, and even if the valence
and sea quark masses are different ({\it i.e.}, even if the theory is 
partially quenched).  With three staggered flavors 
and  $\root 4 \of {\rm Det}$'s,
the theory is, I believe, still in the right
sector and has physical chiral parameters. But it is like a theory with
12 sea quarks, each with $1/4$ weight, rather than 3 normal flavors.

In order to extract the physical chiral parameters from an ordinary
3-flavor partially quenched theory, we need the correct functional forms
calculated in partially quenched chiral perturbation theory. 
Similarly, in order to extract the physical chiral parameters from a theory of
3 staggered flavors with $\root 4 \of {\rm Det}$'s, we
need the functional forms calculated in a staggered chiral perturbation theory
(S\cpt). This includes the effect of the ${\cal O}(a^2)$ taste violations.

The starting point of S\cpt\ is the chiral Lagrangian of Lee and Sharpe \cite{LEE-SHARPE},
which is the 
low energy effective theory for a single staggered field, correct to 
${\cal O}(a^2)$.  To apply it to the case of interest, one 
must generalize to 3 flavors
(which turns out to be non-trivial),  calculate relevant quantities at 1-loop,
and the  adjust for the effect of taking $\root 4 \of {\rm Det}$'s.
Student Chris Aubin and I have done this for $m_\pi^2/(2\hat m)$
and $m_K^2/(\hat m + m_s)$ \cite{CB-CHIRAL}. ($\hat m$ is the average $u,d$ quark mass.)
One can fit the MILC data very well with our results.
We are in the process of extending this work to $f_\pi$, $f_K$ and
heavy-light decay constants, as well as allowing for different valence and
sea quark masses.

\bigskip

{\bf Hashimoto:}
Since the computational cost required to simulate dynamical
quarks grows very rapidly as the sea quark mass is decreased,
controlled chiral extrapolation is crucial to obtain
reliable predictions for physical quantities.
Through this short presentation I would like to share our 
experience with chiral extrapolations obtained from the 
unquenched simulation being performed by the JLQCD
collaboration using nonperturbatively $O(a)$-improved
Wilson fermions on a relatively fine lattice, $a\sim$ 0.1~fm.
Further details are presented in a parallel talk
\cite{Hashimoto_parallel}.

The strategy we have in mind when we do the chiral
extrapolation is to use chiral perturbation theory
(\cpt) as a theoretical guide to control the quark mass
dependence of physical quantities.
For this strategy to work one has to push the sea quark mass
as light as possible and test whether the lattice data
are described by the one-loop \cpt\ formula.
(The lowest order \cpt\ prediction usually does not have quark mass
dependence.) 
If so, chiral extrapolation down to the
physical pion mass is justified.

\begin{figure}[t]
  \begin{center}
    \leavevmode
    \includegraphics*[width=6.9cm,clip]{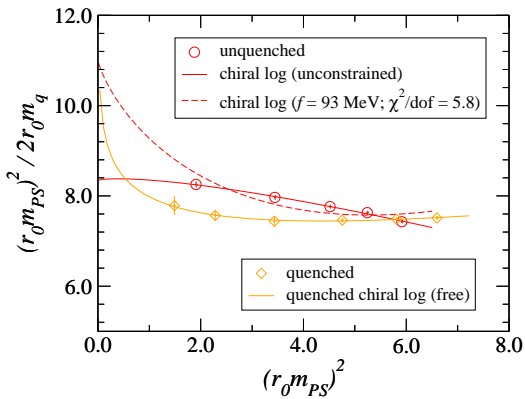}
  \end{center}
  \vspace*{-12mm}
  \caption{
    Test of the one-loop \cpt\ formula from the JLQCD
    collaboration. 
    }
  \label{fig:B_vs_mpi2}
\end{figure}

In full QCD \cpt\ predicts the chiral logarithm with a
definite coefficient depending only on the number of active
flavors, which gives a non-trivial test of the unquenched
lattice simulations.
For example the PCAC relation 
$M_{\mathrm{SS}}^2\simeq2B_0 m_S$
is given as
\begin{eqnarray}
  \label{eq:ChPT_PSmass}
  \lefteqn{
    \frac{M_{\mathrm{SS}}^2}{2B_0 m_S}
    = 1 + \frac{1}{N_f} y_{\mathrm{SS}}\ln y_{\mathrm{SS}}
    } \nonumber\\
  & & 
  + y_{\mathrm{SS}} [
  (2\alpha_8-\alpha_5) + N_f (2\alpha_6-\alpha_4)
  ]
\end{eqnarray}
for $N_f$ flavors of degenerate quarks with a mass $m_S$ and 
$y_{\mathrm{SS}}=2B_0m_S/(4\pi f)^2$.
Similar expression for the pseudoscalar meson decay constant
is 
\begin{equation}
  \label{eq:ChPT_PSdecayconst}
    \frac{f_{SS}}{f}
    = 1 - \frac{N_f}{2} y_{\mathrm{SS}}\ln y_{\mathrm{SS}}
    + \frac{y_{\mathrm{SS}}}{2} [
    \alpha_5 + N_f \alpha_4
    ].
\end{equation}
The coefficient of the chiral log term is fixed, while
the low energy constants $\alpha_i$ are unknown.
Figure~\ref{fig:B_vs_mpi2} shows the comparison of lattice
data with (\ref{eq:ChPT_PSmass}), and it is unfortunately
clear that the lattice result does not reproduce the
characteristic curvature of the chiral logarithm.
The same is true for the pseudoscalar meson
decay constant, and the ratio test using partially
quenched \cpt\ leads to the same conclusion
\cite{Hashimoto_parallel}. 

The most likely reason is that the dynamical quarks in
our simulations are still too heavy.
In fact, the corresponding pseudoscalar meson mass ranges
from 550 to 1,000~MeV, for which we do not naively expect
that \cpt\ works, especially at the high end.
Our analysis of the partially quenched data suggests that
a meson mass as low as 300~MeV is necessary to be
consistent with one-loop \cpt.

Let us now discuss the systematic uncertainty in the
chiral extrapolation.
Since we know that the \cpt\ is valid for small enough quark
masses, the chiral extrapolation has to be consistent with
the one-loop \cpt\ formula at least in the chiral limit.
If we assume that the chiral logarithm dominates only below
a scale $\mu$, a possible model is to take the one-loop \cpt\
formula below $\mu$ while using a conventional polynomial
fitting elsewhere.
Both functions may be connected so that their value
and first derivative match at the scale $\mu$.
The scale $\mu$ is unknown, though we naively expect that
$\mu$ is around 300--500~MeV. 
Therefore, we should consider the dependence on $\mu$ 
in a wider range, say 0--1,000~MeV, as an indication of the
systematic error in the chiral extrapolation.
A plot showing these fitting curves is presented in
\cite{Hashimoto_parallel}.

\begin{figure}[t]
  \begin{center}
    \leavevmode
    \includegraphics*[width=6.5cm,clip]{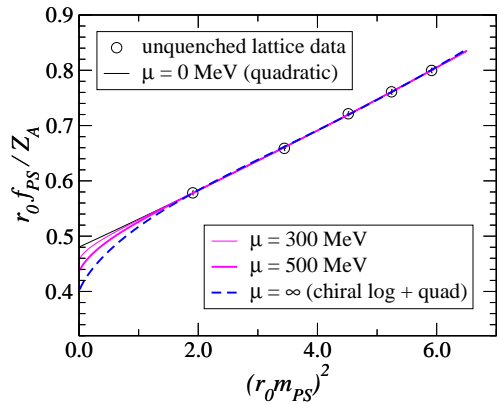}
  \end{center}
  \vspace*{-12mm}
  \caption{
    Uncertainty in the chiral extrapolation of pion decay
    constant. 
    }
  \vspace*{-4mm}
  \label{fig:fpi_vs_mpi2_Adelaide-MIT}
\end{figure}

Another possible functional form is that suggested by the
Adelaide-MIT group \cite{Detmold:2001jb}.
They propose using the one-loop \cpt\ formula calculated
with a hard momentum cutoff $\mu$, which amounts to replacing
the chiral log term
$m_\pi^2\ln(m_\pi^2/\mu^2)$
by 
$m_\pi^2\ln(m_\pi^2/(m_\pi^2+\mu^2))$.
It is a model in the sense that we use it above the cutoff
scale $\mu$.
Fits to the pion decay constant are shown in
Figure~\ref{fig:fpi_vs_mpi2_Adelaide-MIT}.
The fit curves represent the model with $\mu$ = 0, 300, 500,
and $\infty$~MeV.
Since we do not have a solid theory to choose the cutoff
scale $\mu$, the variation of the chiral limit should be
taken as the systematic uncertainty, whose size is of
order of $\pm$10\%.

The large uncertainty associated with the chiral extrapolation
as discussed above has not attracted much attention, partly
because most simulations have been done in the quenched
approximation, for which the chiral behavior of physical
quantities is quite different.
In contrast, in unquenched QCD, confirming the predictions
of \cpt\ gives a non-trivial
test of the low energy behavior obtained from lattice calculations.
For pion or kaon physics it is essential to perform the
lattice simulation in a region where \cpt\ is applicable,
since the physics analysis often relies on \cpt.

State-of-the-art unquenched lattice simulations using
Wilson-type fermions are still restricted to the large
sea quark mass region $\geq m_s/2$, for which we do not find
an indication of the one-loop chiral logarithm.
This means that there could be a sizable systematic
uncertainty in the chiral extrapolation.
I have discussed the example of the pion decay constant; a similar
analysis is underway for the heavy-light decay constants and
light quark masses.

\bigskip

{\bf Pallante:}
Chiral extrapolation of weak matrix elements
and in particular kaon matrix elements (i.e. $B_K$, $K\to 2\pi$ decays, 
semileptonic kaon decays) is a very delicate issue.
One of the most difficult tasks still remains the calculation of 
$\langle\pi\pi |O_W|K\rangle$ matrix elements,
 where $O_W$ is a weak four-quark
effective operator at scales $\mu < M_W$, typically $\mu\sim m_c$.

Since elastic (soft) final state interactions (FSI) of the two pions 
are large especially in the total isospin zero channel (see \cite{PP} 
and refs. therein),
it is mandatory to overcome the {\em Maiani-Testa no-go} theorem \cite{MT}
and to include the bulk of FSI effects directly in the lattice measurement 
of kaon matrix elements, while keeping under control residual corrections
 through the use of Chiral Perturbation Theory ($\chi$PT).

A considerable step forward in this respect has been made in 
refs.~\cite{LL,LMST}, where it has been shown that the physical matrix element
can be extracted from the measurement of an Euclidean correlation function
at {\em finite volume}. The {\em finite volume} matrix element is converted 
to the infinite volume one via a multiplicative {\em universal} factor 
\cite{LL} (denoted as LL factor in the following),
i.e. only depending on the quantum numbers of the $\pi\pi$ final state.

There are three main reasons why $\chi$PT is needed, at least up to 
next-to-leading order (NLO), in extracting the physical 
$\langle\pi\pi |O_W|K\rangle$ matrix element from a lattice 
Euclidean correlation function:
1) Lattice simulations are presently performed at unphysical values of 
light quark masses, so that $\chi$PT is needed to parameterize mass 
dependences 
and perform the extrapolation to their physical value,
provided it is applicable at the values of quark masses used on the lattice.
2) Lattice simulations may be done with unphysical choices of the kinematics, 
simpler than the physical one, and again $\chi$PT is needed for the 
extrapolation.
3) $\chi$PT is an appropriate tool to monitor in a perturbative manner the 
size of 
{\em systematic errors} due to a) (partial) quenching, b) finite volume and 
c) non-zero lattice spacing \cite{Shoresh}.

It is also important to note that 
 the possibility of computing lattice matrix elements
with choices of momenta and masses different from the physical ones is a very 
powerful method, once we want to determine the low-energy constants 
(LEC) which appear in the chiral expansion of observables at NLO.
By varying momenta, and masses, we can increase the number of linear 
combinations of LEC that can be extracted from a lattice computation.

A possible strategy for the {\em direct} measurement of 
$\langle\pi\pi \left|O_W\right|K\rangle$ matrix elements has been formulated 
in 
ref.~\cite{dt32fv}, specifically for the $\Delta I=3/2$ case (see also 
ref.~\cite{david} at this conference). This strategy is general,
and can be applied also to the $\Delta I=1/2$ case.
It is as follows: 
1) evaluate the Euclidean correlation function 
$C_{\vec{q}_1\vec{q}_2}= \langle 0 \left|\pi_{\vec{q}_1}(t_1)\pi_{\vec{q}_2}
(t_2)O_W(0)K_{\vec{0}}(t_K)\right|0\rangle$ at {\em fixed physics}, 
i.e. at fixed two-pion total energy at finite volume;
2) divide by the appropriate source (sink) correlation functions
at finite volume. This step produces the finite volume matrix element 
$\left|{}_V\langle\pi\pi \left|O_W(0)\right|K\rangle{}_V\right|$, and
3) multiply it by the universal LL factor to get the infinite 
volume amplitude: 
$\left|\langle\pi\pi \left|O_W(0)\right|K\rangle\right|^2 = LL\times  
\left|{}_V\langle\pi\pi \left|O_W(0)\right|K\rangle{}_V\right|^2$. 
4) If not able to apply the procedure 1) to 3) directly for the physical 
kinematics, then apply the procedure for an alternative choice of 
the kinematics that is sufficient to fully determine the physical amplitude 
at NLO in $\chi$PT.
Two such choices for the $\Delta I=3/2$ case are the SPQR kinematics 
\cite{SPQR}, where one of the two pions carries a non-zero three-momentum, 
and the strategy proposed in ref.~\cite{soni}, using the combined measurements
of $K\to\pi\pi$ at $m_K=2m_\pi$ and $m_K=m_\pi$, $K\to\pi$ at 
non-zero momentum, $K\to 0$
and $K^0-\bar{K}^0$ transition amplitudes. The second strategy is also 
sufficient for the $\Delta I=1/2$ case, while the SPQR kinematics for 
$\Delta I=1/2$ is under investigation. Also, the LL factor derived in 
\cite{LL} is only applicable to the center-of-mass frame, while its 
generalization to a moving frame has not yet been derived 
(see ref.~\cite{dt32fv} for a discussion).

Unfortunately, most realistic lattice simulations are still  performed in 
the quenched approximation or, at best,
in the partially quenched approximation with 
two or three dynamical (sea) flavors.
The loss of unitarity due to (partial) quenching of  $SU(3)_L\times SU(3)_R$
chiral group has dramatic consequences in the $I=0$ channel of $K\to \pi\pi$ 
amplitudes \cite{all0,scalar}. 
Loss of unitarity implies the failure 
of Watson's theorem and L\"uscher's quantization condition. 
As a consequence, the FSI phase extracted from the quenched weak amplitude 
is no longer universal (i.e. it may also depend on the weak operator $O_W$) 
and finite volume 
corrections of quenched  weak matrix elements among physical states 
are not universal
(i.e. the universality of the LL factor does not work as in the full 
theory) \cite{scalar}.

The reason why the $I=0$ case is a peculiar one is that the rescattering 
diagram of the two final state 
pions (the one producing the phase of the amplitude) is modified by 
(partial) quenching already at one loop in $\chi$PT. 
This is not the case for $I=2$ however, where the rescattering diagram is 
unaffected by quenching at least to one loop in $\chi$PT. This guarantees
the applicability of the {\em direct} strategy to the $I=2$ channel 
also in the quenched 
approximation, at least up to one loop in the chiral expansion.

Another consequence of (partial) quenching is the contamination of QCD-LR 
penguin operators, like $Q_6$, by new non-singlet operators \cite{GP_penguin}
which appear at 
leading order in the chiral expansion (i.e. order $p^0$, even enhanced 
respect to the order $p^2$ singlet operator). 
This contamination does not 
affect $\Delta I=3/2$ transitions, being pure $\Delta I=1/2$.

Given the above picture, a few conclusions can be drawn. At present,
$\chi$PT plays a crucial role in the extrapolation of lattice weak 
matrix elements
to their physical value, or to the chiral limit. 
However, the applicability of $\chi$PT at the lowest orders 
(typically up to NLO) in the extrapolation 
procedure is guaranteed only for sufficiently light
values of lattice meson masses.
This means that one should work in a region of quark masses 
sufficiently far below the first relevant resonance.
The situation can be further complicated by the presence of FSI effects,
especially in the $I=0$ channel. These effects can be either analytically 
resummed \cite{PP} or the bulk of them be directly included  into the finite 
volume lattice matrix element.
Most critical appears the situation in the presence of quenching, due to the 
lack of unitarity. 
For $\Delta I=3/2$ matrix elements, strategies 
proposed for a {\em direct} measurement with 
unquenched simulations can still be used in a quenched simulation at least 
up to NLO in the chiral expansion. This is no longer true for $\Delta I=1/2$
matrix elements.
In this case, quenching and partial quenching affect {\em universal} 
properties of the weak amplitude already at one loop in $\chi$PT, and in 
addition produce a severe contamination of QCD-LR penguin operators with
new non-singlet operators. 
However, those problems disappear in the partially quenched case with
$N_{sea}=N_{valence}$ and $m_{sea}=m_{valence}$, where partially quenched 
correlation functions reproduce those of full 
QCD\cite{CB-MG-PQQCD,SHARPE-SHORESH}.

\bigskip

{\bf Leinweber:}
Until recently, it was difficult to establish the range of quark
masses that can be studied using chiral perturbation theory ($\chi$PT)
\cite{Hatsuda:tt}.  Now, with the advent of lattice QCD simulation
results approaching the light quark mass regime, considerable light
has been shed on this important question
\cite{Leinweber:1998ej,Leinweber:1999ig,Detmold:2001jb,%
Hackett-Jones:2000qk,Leinweber:1999nf,Hackett-Jones:2000js,TonyConf}.
It is now apparent that current leading-edge dynamical-fermion
lattice-QCD simulation results lie well outside the applicable range
of traditional dimensionally-regulated (dim-reg) $\chi$PT in the
baryon sector.

The approach of the Adelaide group is to incorporate the known or
observed heavy quark behavior of the observable in question and the
known nonanalytic behavior provided by $\chi$PT within a single
functional form which interpolates between these two regimes.  The
introduction of a finite-range regulator designed to describe the
finite size of the source of the meson cloud of hadrons achieves this
result.  The properties of the meson-cloud source are parameterized
and the values of the parameters are constrained by lattice QCD
simulation results.  Without such techniques, one cannot connect
experiment and current dynamical-fermion lattice-QCD simulation
results for baryonic observables.

The use of a finite-range regulator might be confused with modeling.
However, it is already established that $\chi$PT can be formulated
model independently using finite-range regulators such as a dipole
\cite{Donoghue:1998bs}.  The coefficients of leading-nonanalytic (LNA)
terms are model independent and unaffected by the choice of regulation
scheme.  The explicit dependence on the finite-range regulation
parameter is absorbed into renormalized coefficients of the chiral
Lagrangian.

The shape of the regulator is irrelevant to the formulation of
$\chi$PT.  However, current lattice simulation results encourage us to
look for an efficient formulation which maximizes the applicable
pion-mass range accessed via one- or two-loop order.  
An optimal regulator (perhaps motivated by phenomenology) will
effectively re-sum the chiral expansion encapsulating the physics in
the first few terms of the expansion.  The approach is systematically
improved by simply going to higher order in the chiral expansion.  Our
experience with dipole and monopole vertex regulators indicates that
the shape of the regulator has little effect on the extrapolated
results, provided lattice QCD simulation results are used to constrain
the optimal regulator parameter on an observable-by-observable basis
\cite{Leinweber:2001ac}.

In order to correctly describe QCD, the coefficients of nonanalytic
terms must be fixed to their known model-independent values.  This
practice differs from current common practice within our field where
these coefficients are demoted to fit parameters and optimized using
lattice simulation results which lie well beyond the applicable range
of traditional dim-reg $\chi$PT.  The failure of the approach is
reflected in fit parameters which differ from the established values
of $\chi$PT by an order of magnitude \cite{Leinweber:1999ig} spoiling
associated predictions \cite{Leinweber:2001ac,Leinweber:2000sa}.

\begin{figure}[t]
\centerline{\psfig{file=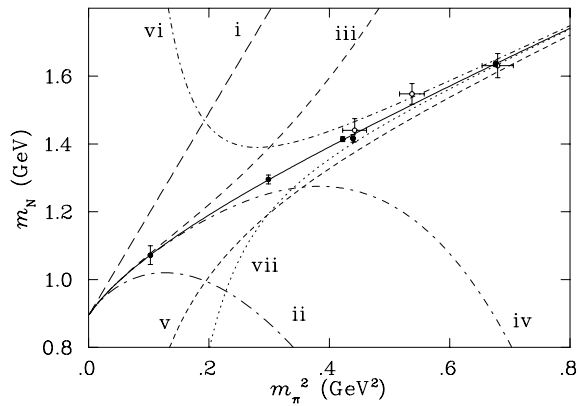,width=\hsize,angle=90}}
\vspace*{-1.0cm}
\caption{The finite-range regulated expansion of the nucleon mass
(solid curve) and its perturbative chiral (curves (i) through (iv) and
heavy-quark (curves (v) through (vii)) expansions
\protect\cite{swright}.  Details of each curve may be found in the
text.
\vspace{-12pt}}
\label{slide1}
\end{figure}

I will focus on the extrapolation of the nucleon mass as it
encompasses the important features which led to subsequent
developments
\cite{Leinweber:1998ej,Detmold:2001jb,Hackett-Jones:2000qk,%
Leinweber:1999nf,Hackett-Jones:2000js} required to extrapolate today's
lattice QCD results.  Figure \ref{slide1} displays the results of a
finite-range chiral expansion \cite{swright} of the nucleon mass
(solid curve) constrained by dynamical-fermion simulation results from
UKQCD \cite{Allton:1998gi} (open symbols) and CP-PACS
\cite{Aoki:1999ff} (closed symbols).  The expression for the nucleon
mass 
\begin{eqnarray}
M_N &=& c_0 + c_2 \, m_\pi^2 -
{{3 \, g_A^2 \over 32 \, \pi \, f_\pi^2}} \times \label{NPexp} \\
&& {2 \over \pi} \,
\left [
m_\pi^3 \; {\rm arctan}\, \left ( {{\Lambda \over m_\pi}} \right ) +
{\Lambda^3 \over 3} - \Lambda\, m_\pi^2 
\right ] \, , \nonumber
\end{eqnarray}
arises from the one-loop pion-nucleon self-energy of the nucleon, with
the momentum integral regulated by a sharp cutoff.  The lattice
simulation results constrain the optimal regulator parameter,
$\Lambda$, to 620 MeV.  
Of course it is desirable to use more realistic regulators such as a
dipole form when keeping only one-loop terms of the chiral expansion.

For small $m_\pi$ the standard LNA behavior of $\chi$PT is obtained
with the correct coefficient.  For large $m_\pi$, the arctangent tends
to zero and suppresses the nonanalytic behavior in accord with the
large quark masses involved.  The scale of the regulator $\Lambda$ has
a natural explanation as the scale at which the pion Compton wave
length emerges from the hadronic interior.  It is the scale below
which the neglected extended structure of the effective fields becomes
benign.

The valid regime of the truncated expansion of $\chi$PT is the regime
in which the choice of regulator has no significant impact.  To gain
further insight into the validity of the truncated expansion of
traditional dim-reg $\chi$PT, one can perform a power series expansion
of the arctangent in terms of $m_\pi/\Lambda$ and keep terms only to a
given power \cite{swright}.  The dim-reg expansion of (\ref{NPexp})
for small $m_\pi/\Lambda$ is provided by curves (i) through (iv) in
Fig.~\ref{slide1}.  Curve (i) contains terms to order $m_\pi^2$, and
(ii) to order $m_\pi^3$.  This is the correct implementation of the
LNA behavior of $\chi$PT.  The behavior dramatically contrasts the
common but erroneous approach discussed above.  The applicable range
of traditional dim-reg $\chi$PT to LNA order for the nucleon mass is
merely $m_\pi < 200$ MeV.  Incorporation of the next analytic
$m_\pi^4$ term extends this range to 400 MeV.  Curve (iv) illustrates
the effect of including the $m_\pi^6$ term of the expansion.

Within the range $m_\pi^2 < 0.15\ {\rm GeV}^2$, dim-reg $\chi$PT
requires three analytic-term coefficients, $c_0$, $c_2$ and $c_4$ (the
coefficient of $m_\pi^4$), to be constrained by lattice QCD simulation
results.  
The Adelaide approach optimizes $c_0$, $c_2$ and $\Lambda$ in place of
$c_4$.  Tuning the regulator parameter is not modeling.  Instead,
optimization of the regulator provides the promise of suppressing
$c_4$ and higher-order terms.  One can understand how this
approach works through the consideration of how the regulator models
the physics behind the effective field-theory, but such descriptions
do not undermine the rigorous nature of the effective field theory.

While the Adelaide approach of (\ref{NPexp}) {\it is} $\chi$PT, it is
the current state of lattice QCD simulation results that demand the
parameters of the chiral expansion be determined in other ways.  The
extension to generalized Pad\'e approximates
\cite{Leinweber:1998ej,Hackett-Jones:2000qk,Leinweber:1999nf,%
Hackett-Jones:2000js}, modifications of log arguments
\cite{Detmold:2001jb} and meson-source parameterizations
\cite{Leinweber:1999ig,Leinweber:2001ac,Leinweber:2000sa,%
Young:2002cj} are methods to constrain the chiral parameters with
today's existing lattice QCD simulation results.

Traditional dim-reg $\chi$PT to one loop knows nothing about
the extended nature of the meson cloud source.  
As there is no other mechanism to incorporate this physics,
the expansion fails catastrophically if it is used beyond the
applicable range.  Moreover, convergence of the dim-reg expansion is
slow as large errors associated with short-distance physics in loop
integrals (not suppressed in dim-reg $\chi$PT) must be removed by
equally large analytic terms.  These points are made obvious by
examining the predictions of the power series expansions (curves (i)
through (iv)) of Fig.~\ref{slide1} at $m_\pi^2 = 0.3\ {\rm GeV}^2$.
Curve (ii) incorporating terms to $m_\pi^3$ is particularly amusing.

In contrast, the optimal finite-range regulation of the Adelaide Group
provides an additional mechanism for incorporating finite-size
meson-cloud effects beyond that contained explicitly in the leading
order terms of the dim-reg expansion.  The finite-size regulator
effectively re-sums the chiral expansion, suppressing higher-order
terms and providing improved convergence.  The net effect is that a
catastrophic failure of the chiral expansion is circumvented and a
smooth transition to the established heavy quark behavior is made.

It is time for those advocating standard chiral expansions to use them
with the established model-independent coefficients and in a regime
void of catastrophic failures; a regime that can be extended using
finite-size regulators.  The approach of the Adelaide Group provides a
mechanism for confidently achieving these goals with the
cautious conservatism vital for the future credibility of our field.

\bigskip

{\bf Lepage:}
This is a remarkable time in the history of lattice QCD. For the 
first time we appear to have an affordable procedure for almost 
realistic unquenching. Improved staggered quarks are so efficient 
that the MILC collaboration has already produced thousands of 
configurations with small lattice spacings and three flavors of light 
quark: one at the strange quark mass, and the other two at masses of 
order $1/5$ or $1/7$ or less of the strange quark mass. For the first 
time we can envisage a broad range of phenomenologically relevant 
lattice calculations, in such areas as $B$~physics and hadronic 
structure, that are precise to within a few percent {\em and} that 
must agree with experiment.

Chiral extrapolations are likely to be one of the largest sources of 
systematic error in such high-precision work. The MILC collaboration 
is already working at much smaller light-quark masses than have been 
typical in the field; there is little doubt that these masses are 
small enough for a viable chiral perturbation theory. And partial 
quenching provides a powerful tool for determining the needed chiral 
parameters. Such a systematic approach is essential for high 
precision.

As discussed by Claude Bernard, the most significant complication in 
the chiral properties of improved staggered quarks comes from their 
``taste-changing'' interactions. Crudely speaking these generate a 
non-zero effective quark mass, proportional to $a^2$, even for zero 
bare quark mass. This effect is perturbative in QCD, and can be 
removed by modifying the quark action. It can also be measured 
directly in simulations; we should  know shortly how significant it 
is for typical lattice spacings.

An important aspect of high-precision lattice QCD is choosing 
appropriate targets. High-precision work in the near future will 
focus on stable or nearly stable hadrons.
It will be much harder to achieve errors smaller 
than 10--20\% for processes that involve unstable hadrons such as the 
$\rho$ or $K^*$. One might try to extrapolate through the decay 
threshold, but thresholds are intrinsically nonanalytic and so 
extrapolation is very unreliable. Hadrons very near to thresholds, 
such as the $\phi$ or the $\psi^\prime$, may be more accessible, but 
even these will be unusually sensitive to the light-quark mass since 
this affects the location of the threshold.

Such considerations will dictate which simulations we do and how we 
do them. Consider, for example, how we set the physics parameters
in a simulation. The $1S-1P$ splittings in the $\psi$ or $\Upsilon$ 
systems are ideal for determining the lattice spacing. 
The hadrons involved are well below the $D$-$\overline{D}$ and 
$B$-$\overline{B}$ thresholds. They have no valence $u$ and 
$d$~quarks, and couple 100 or 1000 times more weakly to $\pi$s than 
ordinary mesons. This means these splittings are almost completely 
insensitive to light-quark masses (once these are small enough). 
Finally, and somewhat surprisingly, the splittings are almost 
completely insensitive the $c$~and $b$~quark masses as well. To 
a pretty good approximation, the only thing these splittings depend 
upon is~$a^{-1}$. Bad choices for setting~$a^{-1}$ would be the 
$\rho$~mass or even the $\psi'$-$\psi$ splitting, since the $\psi'$ 
is only 40\,MeV away from a threshold.

Another example concerns setting the strange quark mass. Obvious 
choices for this are the splittings $2 M(B_s) - M(\Upsilon)$ and $2 
M(D_s)-M(\psi)$.
These involve no valence $u$ and $d$ quarks, and so require much less 
chiral extrapolation than say $M(K)$. Also they are, by design, 
approximately independent of the heavy quark masses as well. And each 
of the hadrons is far from thresholds. To a pretty good approximation, 
these splittings depend only upon~$m_s$.

The CLEO-c experiment presents a particularly exciting opportunity 
for lattice QCD, as discussed by Rich Galik at this meeting. Within 
about 18~months CLEO-c will start to release few percent accurate 
results for  $f_D$, $D\to\pi l\nu$, $f_{D_s}$\,\ldots. A challenge for 
lattice QCD is to {\em predict} these results with comparable 
precision. This would provide much needed credibility for 
high-precision lattice QCD, substantially increasing its impact on 
heavy-quark physics generally. It would also be a most fitting way to 
celebrate lattice QCD's $30^\mathrm{th}$ anniversary.

\bigskip

{\bf Wittig:}
Further to the issues discussed in my plenary talk \cite{HW_plenary},
I would like to focus on two questions, namely
\begin{itemize}
\item How can we gain information on physical quantities in the
      {\it most reliable} way?
\item How can we check the validity of $\chi$PT?
\end{itemize}
As an example let me come back to the masses of the light
quarks. Their absolute values are not accessible in $\chi$PT, but
quark mass ratios have been determined at NLO, using values for the
low-energy constants (LEC's) that were estimated from phenomenology in
conjunction with theoretical assumptions \cite{Leutwyler96}. The
results are
\begin{eqnarray}
   & &\frac{m_{\rm u}}{m_{\rm d}}=0.553\pm0.043,\quad 
      \frac{m_{\rm s}}{m_{\rm d}}=18.9\pm0.8 \nonumber \\
   & &\frac{m_{\rm s}}{\widehat{m}}=24.4\pm1.4, \quad
      \widehat{m}=\textstyle\frac{1}{2}(m_{\rm{u}}+m_{\rm{d}}).
      \label{eq_qmratios}
\end{eqnarray}
Individual values can thus be obtained if one succeeds in computing
the absolute normalization in a lattice simulation. The most easily
accessible quark mass on the lattice is surely $m_{\rm s}$, for which
an extrapolation to the chiral regime is not required
\cite{ALPHA_ms}. The combination of the lattice estimate for
$m_{\rm{s}}$ with the ratios in eq.\,(\ref{eq_qmratios}) then yields
the values of $m_{\rm{u}},\,m_{\rm{d}}$ without chiral extrapolations
of lattice data.

This is a reliable procedure, provided that the theoretical
assumptions, which are used to determine some of the LEC's that are
needed for the results in (\ref{eq_qmratios}), are justified. Whether
this is the case can be studied in lattice QCD, either by computing
ratios like $m_{\rm s}/\widehat{m}$ or $m_{\rm{s}}/m_{\rm{d}}$
directly on the lattice, or by determining the LEC's themselves in a
simulation. The apparent advantage of the latter is that only
moderately light quark masses are required. Furthermore, it is
difficult---though not impossible---to distinguish between
$m_{\rm{u}}$ and $m_{\rm{d}}$ in lattice simulations.

Can we trust the current lattice estimates for the low-energy
constants? ALPHA and UKQCD \cite{ALPHA_ChPT,UKQCD_mup0} have extracted
them by studying the quark mass behaviour in the range
\begin{equation}
  m_{\rm{s}}/2\;\lesssim\;m\;\lesssim\;m_{\rm{s}}
\end{equation}
In order to check whether lattice estimates for the low-energy
constants make sense phenomenologically, we can use the results for
$\alpha_5$ to predict the ratio of decay constants $F_{\rm{K}}/F_\pi$,
whose experimental value is $1.22\pm0.01$.

UKQCD have simulated $n_{\rm{f}}=2$ flavours of dynamical quarks. For
the sake of argument, let us assume that the quark mass dependence
is not significantly different in the physical 3-flavour case. The
data can then be fitted using the expressions in partially quenched
$\chi$PT for $n_{\rm{f}}=3$. In this way one obtains
\begin{equation}
 ``{\alpha_5^{(3)}}`` = 0.98\pm0.09\pm0.25,
\end{equation}
where the first error is statistical, the second is systematic, and
the inverted commas remind us that this is not really the 3-flavour
case. After inserting this estimate into the expression for
$F_{\rm{K}}/F_\pi$ in ``full'' QCD \cite{GaLeu85} one obtains
\begin{equation}
 F_{\rm{K}}/{F_\pi} = 1.247\pm0.011\pm0.020,
\label{eq_FKFpi}
\end{equation}
which is consistent with the experimental result. This is an
indication that the quark mass dependence of pseudoscalar decay
constants in the physical 3-flavour case is not substantially
different from the simulated 2-flavour theory.

It is interesting to note that the estimate in eq.\,(\ref{eq_FKFpi})
decreases by 15\% if the chiral logs are neglected in the expression
for $F_{\rm{K}}/{F_\pi}$, i.e.
\begin{equation}
 F_{\rm{K}}/{F_\pi} = 1.080\pm0.007\pm0.021.
\end{equation}
This example then demonstrates that the inclusion of chiral logarithms
can significantly alter predictions for SU(3)-flavour breaking ratios
such as $F_{\rm K}/F_\pi$. This observation was also made recently by
Kronfeld \& Ryan in the context of the corresponding ratio for B-meson
decay constants, i.e. $F_{\rm{B}_s}/F_{\rm{B}_d}$
\cite{KRONFELD}. Unlike the situation for $F_{\rm K}/F_\pi$,
however, there is no experimental value to compare with.

To summarize: these examples serve to show that estimates for light
quark masses can be obtained in a reliable way by combining lattice
simulations with $\chi$PT, whose strengths and weaknesses are largely
complementary. In order to arrive at mass values for the up- and down
quarks, the ``indirect'' approach via the determination of low-energy
constants offers clear advantages over attempts to compute these
masses directly in simulations.

\section{Responses}
{\bf Bernard:}
The aim of Lattice QCD (LQCD) 
is to predict numbers in a controlled way.
The problem introduced by the Adelaide approach
is that nearly any functional form will fit 
across the linear portion of the data but model-dependent constants 
are being introduced that can change the extrapolated answer
by an unknown amount. Although changes in the 
chiral regulator are ordinarily thought of as
harmless, since they can be absorbed into changes in the analytic terms,
that is not true when theory is used to fit data in the regime above the
cutoff, in the linear regime.  The detailed form of the cutoff is then
important, and there is no universality.

We need to {\em fit} the chiral logs in a 
controlled manner.  Indeed, we can now do this with the improved
staggered fermion data, which extends down to $m_\pi/m_\rho\approx 0.35$,
as long as we use the appropriate chiral Lagrangian.

Another approach that should be pursued 
is using chirally improved or fixed-point fermions for
valence quarks on dynamical configurations generated with
improved staggered quarks. This would provide important tests
of staggered results.

\medskip

{\bf Hashimoto:}
First, I agree with Claude Bernard about the importance
of distinguishing rigorous 
lattice calculations from those with model dependence.  
Nevertheless, I think that models are
a useful way of estimating systematic uncertainties.

My second remark concerns the extraction of low energy constants
(LEC) in the chiral Lagrangian from fits to lattice data.
To do this, we must first check that \cpt\ fits the lattice data.
When the sea quark mass is too large,
the \cpt\ formulae will not work.  
We find that they do not work for JLQCD's data, 
and thus do not quote results for LEC.
The situation might, however, be better with the staggered
fermion data.

Third, staggered fermions have the advantage of
allowing one to push sea-quark masses closer to 
the chiral limit, but Wilson fermions are useful for their 
simplicity and should be used as a cross-check.

\medskip

{\bf Pallante:}
Present chiral extrapolations for weak matrix element calculations
use $m_\pi > 400\;$MeV, and this is too high to trust \cpt.
We need to bring the mass down to $300\;$MeV.
It may be that to work reliably in this regime requires
chirally symmetric fermions. In this regard, the approach mentioned
by Giusti~\cite{Giusti02} is very interesting: matching results
for correlation functions in small volumes to the predictions of
chiral perturbation theory in order to calculate LECs.
The use of small volumes may allow one to work with dynamical
chirally symmetric fermions. At the very least,
this should be pursued as a complementary approach to allow comparisons.

I also agree with Claude Bernard about the dangers of modeling.
I think we have enough theoretical tools
given the lightness of light quarks
and the heavy quark expansion that we can 
control systematic errors, which we cannot do in a model.

\medskip

{\bf Leinweber:}
We in Adelaide are not interested in modeling, either.  If we were
modeling, we would fix the value of the regulator parameter $\Lambda$
from phenomenology.  Instead, we determine it, quantity by quantity,
by fitting lattice data in the region where the regulator sets in.
Our aim is to provide a simple analytic {\em parameterization},
incorporating known physics.  Systematic errors can be estimated by
varying the parameterization.  For example, our studies of the $\rho$
meson indicate a systematic error after chiral extrapolation of about
50 MeV.

The lattice community needs to do better than making linear fits just
because the data looks linear.  We {\em know} that there is
nonanalytic behavior at small $m_q$.  Ideally we should calculate at
much smaller $m_q$ where we can use any regulator including
dimensionally regulated \cpt, but until we can do this we need
alternative parameterizations which extend to higher $m_q$.

Finally, I would like to advocate setting the scale using the static
quark potential and not $m_\rho$.  The former is insensitive to light
quark masses, whereas the latter clearly is.

\medskip

{\bf Lepage:}
First, let me note that if we use the potential to set the scale, 
then we should use $r_0 \approx 0.45\;$fm and not $0.5\;$fm as is usually done. The 
traditional value for $r_0$ comes from models, not from rigorous calculations.
We can, however, infer the correct value from other determinations of the lattice
spacing.

Second, let me address the issue of modeling versus using \cpt. Much of what
the Adelaide group does can be interpreted as an implementation of \cpt\ 
that uses a momentum cut-off, rather than dimensional regularization, to
control ultraviolet divergences. Momentum cutoffs, with $\Lambda\sim500\;$MeV,
have been quite useful in applications of \cpt\ to low-energy nuclear 
physics. Typically such cutoffs make it easier to guess the approximate 
sizes of coupling constants that haven't been determined yet.
It would be quite interesting for the Adelaide group or someone else
to explore whether momentum cutoffs lead to benefits in non-nuclear problems.
The use of a momentum cut-off does not, however, extend the reach of chiral 
perturbation theory to higher energies; ultimately the physics is the same, 
no matter what the UV regulator.  I see no problem with the 
Adelaide approach in so far as it is equivalent to \cpt\ with
a momentum cutoff, but this entails a more systematic approach to
the enumeration and setting of parameters.

Finally, I would like to reemphasize the importance of using
small quark masses, and the significance of the fact that
MILC simulations are entering this regime.
This is a new world---one in which we can control all systematic errors.

\medskip

{\bf Wittig:}
Let me first comment on the ``catastrophic failure'' of \cpt\ 
when extended too far observed by JLQCD.  
UKQCD does not see such a failure, and it is important
that the two groups discuss this point and attempt common
methods of analysis.

Concerning models, let me reiterate that I think 
modeling is a dangerous path to follow.
 Models are usually based on one particular mechanism. It is then
 unclear to what extent they are able to capture this aspect at the
 quantitative level, and whether they are general enough to describe
 other related phenomena correctly.
One particular concern is whether a given model can be falsified.
Is it possible to choose the parameters to make 
the results come out correct for some quantities,
but wrong for others?

\section{General Discussion}

{\bf Stamatescu:}
I have heard advocates of different fermion actions: staggered,
Wilson and others.
I was hoping to hear more than simple advocacy,
and think it would be useful to have a comparison of 
the uses of each type of fermion.

\smallskip

{\bf Shoresh:}
Concerning the need to take the
fourth-root of the determinant when using staggered fermions
it has been stated that there is no evidence that it is wrong, 
but that there is no proof that it is correct.
This seems to come under the heading of uncontrolled errors.
Are any of the panelists uncomfortable with staggered fermions?

Pursuing this point, let me note
that chirally symmetric fermions can be 
simulated for lower quark masses and this has 
been done in quenched simulations.  What is the feasibility of 
doing dynamical simulations with, say, overlap fermions?

\smallskip

{\bf Lepage:}
The reason I am pushing
staggered fermions is that these are the only 
calculations that have a chance to be ready within 18 months.  
The others don't have that chance.

\smallskip

{\bf Bernard:}
The issue of taking roots of the determinant
is certainly an important concern for those of us using
staggered fermions.
One way to study this issue is to compare
results from simulations to the theoretical predictions of
chiral perturbation theory including $O(a^2)$ ``taste'' violations.
If successful, this will show not only that staggered fermions have the
correct chiral behavior in the continuum limit but also that
we understand and can control the approach to that limit.
That should go a long way towards reassuring those who are
skeptical of staggered quarks.

\smallskip

{\bf Lepage:}
Precision calculations can also provide an important check.
Once we have half-dozen quantities calculated at the few~\% level 
and agreeing with experiment, it will increase our confidence.

\smallskip

{\bf Golterman:}
I would like to emphasize the importance of the
issue of whether the strange quark is light enough to be in the 
chiral regime. This is very important for
present calculations of kaon weak matrix elements, 
which all rely on \cpt, and actually extract LECs, rather than
physical decay amplitudes.

I am supportive of the use of improved staggered fermions.
We need unquenched results for phenomenological applications.
I am concerned that the numbers coming out of the Lattice Data Group
working groups will be coming primarily from quenched simulations.

\smallskip

{\bf Rajagopal:}
It is possible to generalize the first of the questions posed 
by Steve Sharpe about how small is small enough.  In thermodynamics 
with 2 quarks there is a second-order phase transition
as the mass goes to zero.  
When $m_q$ is small but non-zero there is a well-defined 
scaling function that can be used to gauge how small an 
$m_q$ is small enough.  In this context,
as in the context of Sharpe's question as posed,
it may turn out that small enough means pion masses of order
or smaller than in nature.  Can these two ways of
gauging what is small enough be related?

I would also like to hear the reaction to my take on the
Adelaide approach.
If a calculation of an observable is linearly 
extrapolated and it misses, what can I learn from this?
I think that a model can make plausible that QCD is not wrong.  
I hear Derek Leinweber fighting the urge to use linear fits where 
we know that the data should not be linear.  But, in order
to calculate an observable quantitatively from QCD, say at the 
few percent level with controlled errors,
we must have lattice data, not a model.  The value
of models is that they can yield qualitative understanding, for
example of what physics is being missed by linear extrapolation. 

\smallskip

{\bf Leinweber:}
I agree that to get an answer at the 1\% level, we need new lattice
results at light quark masses.  But we shouldn't throw out the
parameterization of the regulator.  I encourage everyone here to do
the extrapolations with a variety of regulator parameterizations and
verify the uncertainties for themselves.  We do need more light-quark
lattice results, but I think that we can use the Adelaide approach
now to obtain results at the 5\% systematic uncertainty level.

\smallskip

{\bf Giusti:}
With regard to Lepage's comments on competing with CLEO, 
let me make the following comments.
First,  
simulations with overlap fermions have developed very quickly,
and are becoming competitive,
and we should not stop these and start up with improved
staggered fermions.

Second, in the last 15-20 years the errors on $f_D$ and $f_B$
have approximately halved.
How can you expect the errors to go down by a factor of five in one year,
which is what is needed to attain the 1-2\% errors you are aiming for?

\smallskip

{\bf Lepage:}
The errors on $f_{D,B}$ would have been reduced by far more than
half had it not been for the $\sim20\%$ uncontrolled systematic error
due to quenching. The quenching errors dominated all others
because decay constants are very sensitive to unquenching. Given
realistic unquenching, with improved staggered quarks, the dominant 
errors now are in the perturbative matching to the continuum, 
and we know how
to remove them (and are doing it). Again, we are in a new world.

As to your first question,
I am not saying that you should stop what you are doing.  
I am telling you what I am doing!

\smallskip

{\bf Mawhinney:}
Let me note that dynamical simulations
with 2 flavors of domain-wall fermions
using an exact algorithm are {\em already underway}.
The parameters are $1/a\approx  2\;$GeV, $m_q\approx m_s/2$,
and a fifth dimension of $L_s=12$.
The residual mass is $m_{\rm res} \approx m_s/20-m_s/10$.
Thus, although domain-wall fermions are certainly numerically
more intensive than staggered, Wilson, etc., 
they are not so far from simulating QCD.

\smallskip

{\bf Wosiek:}
Is the cut-off $\Lambda$ universal?

\smallskip

{\bf Leinweber:}
No.  We fit it separately to each quantity to optimize the regulator
of the truncated chiral expansion.

\smallskip

{\bf Neuberger:}
Staggered fermions on a CP-2 manifold have a continuum limit,
but there is no spin connection on this manifold.
Does that worry you?

\smallskip

{\bf Bernard:}
What really worries me is that I didn't understand anything you just said.

[More serious response, added after Maarten Golterman and Michael Ogilvie
explained the question and the answer to me: On CP-2, 
the connection between staggered fermions and naive fermions is lost,
so that, although the staggered theory does exist,  it has no relation 
to a theory of particles of spin 1/2.  Equivalently, momentum space on CP-2 is
different from what we are used to, so one cannot make the usual
construction a continuum spin 1/2 field out of the staggered field 
at the corners of the Brillouin zone.]

\smallskip

{\bf Soni:}
I want to stress two related points.
First, that we don't necessarily need experimental data to test
our methods---we can compare results using different
discretizations and methods. $B_K$ is a good example of
such cross-checking---the comparison of results obtained using
staggered and chirally symmetric fermions
will provide a detailed test of our methodology.

Second, regarding CLEO-c, I am worried about trying to guide experimental 
efforts, which are enormously costly,
 toward the fantasies of theorists.
I am worried about telling them what to measure based 
on what quantities we are able to calculate.
If you think that staggered 
fermions can calculate quantities so precisely, then why not
go to the Particle Data Book.  
There are quite a lot of quantities that have {\em already
been measured very precisely}, such as the $D_S^*-D_S$ mass difference
(known to $0.3\%$).

\smallskip

{\bf Lepage:}
Even without the lattice, experimentalists should be 
measuring these quantities. They are important to
test heavy quark effective theory, and as inputs into studies of
$B$-physics. The CLEO-c measurements are important to lattice
QCD because they test the right things, not just the spectrum. 
We use a large, complex collection of techniques; we need a large
number and variety of tests in order to calibrate all of our methods
to the level of a few~\%. 
CLEO-c is uniquely useful for such tests because it will 
accurately measure
the $D$~analogues of precisely the quantities most important to 
high-precision $B$~physics. 

We have promised the government that we can do calculations well,
and it is about time that we came through.  If we can't 
calculate $f_B$ to better than 15\%, then why are they paying 
for us to have computer time?  Maybe we will be humiliated by 
CLEO if our predictions fail,
but since when is that a reason not to try?

\smallskip

{\bf Creutz:}
While we can't calculate at the physical mass, I have always 
thought it was fun to play and change the mass.  In particular,
I am  fascinated by the prediction that if one has
odd number of negative mass fermions then CP is spontaneously
violated.  I think it would be cool if we could simulate 
such a theory on the lattice.  But staggered fermions always generate 
fermions in pairs.  Do you have any ideas or comments about 
this?  (This is a subtle criticism of staggered fermions.)

\smallskip

{\bf Lepage:}
Yes. Somehow whenever I am talking to somebody about staggered 
fermions, they always manage to bring up the one situation that 
we are absolutely sure that we cannot solve with staggered 
fermions. 

%

\smallskip

{\bf Brower:}
Back to that fourth root of the determinant.
How does this work if, before one 
takes the root, there is a different mass for each taste.
Surely, one wants the chiral logs to characterize the 
splitting as it occurs on the lattice.

\smallskip

{\bf Bernard:}
This can be done by using
chiral perturbation theory including $O(a^2)$ taste-violating terms
and making the connection between \cpt\ diagrams 
and ``quark-flow diagrams'' \cite{CB-CHIRAL}. One can determine which
of the meson diagrams correspond to $0,1,2,\dots$ virtual quark loops,
and then multiply each diagram by the correct power of $1/4$.
Essentially, one is putting in the fourth root by hand.

\smallskip

{\bf Savage:}
\cpt\ already has a length scale $\Lambda_\chi=4\pi f_\pi$ built
into it. I do not see how we are gaining anything
by introducing the form-factor cut-off $\Lambda$.

\smallskip

{\bf Leinweber}
For baryons with a sharp cut-off, $\Lambda=\Lambda_\chi/2$, and this
factor of two is very important in practice.  In the meson sector
there is some question as to whether one needs to introduce a
finite-range form-factor style regulator.

It is important to remember that this second scale, $4\pi f_\pi$, is
not a regulator scale in traditional \cpt.  With dimensional
regulation, the pion mass sets the scale of physics associated with
loop integrals.  As the pion mass becomes large, short-distance
physics dominates and the effective field theory undergoes a
catastrophic failure.

There was an earlier comment suggesting that the parameterization of
the regulator introduces model-dependent constants that don't go away.
These constants do go away as they may be absorbed into a
renormalization of the chiral Lagrangian coefficients.



\end{document}